\documentclass{article}

\usepackage{graphicx}  
\usepackage{a4wide}

\usepackage{subcaption}

\usepackage{amsmath,amssymb}
\usepackage{cleveref}

\usepackage{authblk}   

\begin{document}

\title{\textbf{Neuromorphic information processing using ultrafast heat dynamics and quench switching of an antiferromagnet}}  
\date{}
\author[1,2]{J. Zubáč\footnote{Email: zubac@fzu.cz}\footnote{These authors contributed equally.}}
\author[2]{M. Surýnek$^{\dagger}$}
\author[1]{K. Olejník}
\author[1,2]{A. Farkaš}
\author[1]{F. Krizek}
\author[2]{L. Nádvorník}
\author[2]{P. Kubaščík}
\author[1,2]{Z. Kašpar}
\author[2]{F. Trojánek}
\author[3]{R. P. Campion}
\author[1]{V. Novák}
\author[2]{P. Němec}
\author[1,3]{T. Jungwirth}
\affil[1]{Institute of Physics of the Czech Academy of Sciences, Prague, Czech Republic}
\affil[2]{Faculty of Mathematics and Physics, Charles University, Prague, Czech Republic}
\affil[3]{School of Physics and Astronomy, University of Nottingham, United Kingdom}

\maketitle  

\begin{abstract}
Solving complex tasks in a modern information-driven society requires novel materials and concepts for energy-efficient hardware. Antiferromagnets offer a promising platform for seeking such approaches due to their exceptional features: low power consumption and possible high integration density are desirable for information storage and processing or applications in unconventional computing. Among antiferromagnets, CuMnAs stands out for atomic-level scalable magnetic textures, analogue multilevel storage capability, and the magnetic state's control by a single electrical or femtosecond laser pulse. 
Using a pair of excitation laser pulses, this work examines synaptic and neuronal functionalities of CuMnAs for information processing, readily incorporating two principles of distinct characteristic timescales. Laser-induced transient heat dynamics at sub-nanosecond times represents the short-term memory and causes resistance switching due to quenching into a magnetically fragmented state. This quench switching, detectable electrically from ultrashort times to hours after writing, reminisces the long-term memory. The versatility of the principles’ combination is demonstrated by operations commonly used in neural networks. Temporal latency coding, fundamental to spiking neural networks, is utilized to encode data from a grayscale image into sub-nanosecond pulse delays. Applying input laser pulses with distinct amplitudes then allows for pulse-pattern recognition. The results open pathways for designing novel computing architectures.

\end{abstract}

\section{Introduction} \label{sec:01_intro}
Recent breakthroughs in artificial intelligence and computing are revolutionizing diverse industries and our daily lives, driving further research in numerous areas of science and technology. Artificial neural networks are routinely used to solve sophisticated problems ranging from large datasets processing and decision-making over language understanding and text generation to object detection, image classification and pattern recognition.
These networks were initially developed as mathematical abstraction of biological neural networks and, similarly to them,  are comprised of nodes called neurons. 
However, artificial neurons retain only some fundamental characteristics of their biological counterparts.  
They possess adjustable synaptic weights and process information by performing the weighted summation of real-valued analogue inputs. 
The resulting sum is passed to a thresholding nonlinear activation function, 
which is essential for the network to approximate a general nonlinear function \cite{Markovic2020Physics}. 
In contrast with biological neurons, simplified artificial neurons are static, lacking any inherent temporal dynamics.

A somewhat different approach from that used in artificial neural networks is employed in
spiking neural networks. These networks enable event-driven information processing and dynamic time-dependent calculations resembling neuro-biological systems called neuromorphic computing. The building blocks of spiking
neural networks are spiking neurons, with a leaky-integrate-and-fire neuron model as a notable example. 
Like artificial neurons, spiking neurons integrate the weighted inputs and produce an output above a threshold. However, the weights are updated through the biologically inspired learning rule called spike-timing-dependent plasticity \cite{Grollier2020Neuromorphic}.
Instead of working with real-valued numbers, spiking neurons use the relative timing of short pulses called spikes.
Similarly, as the photoreceptors in an eye convert the incoming light into trains of spikes with a frequency corresponding to light intensity, a bright pixel of a digital image can be represented by higher spiking rates than a darker one. This representation of inputs is called rate encoding \cite{Eshraghian2023Training}.
Another timing-based encoding mechanism is termed temporal latency coding. 
The mechanism exploits the delay between individual spikes rather than their frequency. The encoding takes advantage of sparsely distributed spikes, i. e. only two spikes are, in principle, needed to encode a single data point. Benefiting from spike sparsity, it is particularly convenient for efficient data representation. 

For practical employment of these biologically inspired concepts in computing, specifics of particular implementations have to be considered.  Artificial neural networks, used for solving difficult problems of artificial intelligence, owed their success to the increasing computational performance of traditional computers. \cite{amodei2018ai}.
However, increasing energy costs and adverse environmental impacts when solving gradually more complex tasks \cite{Sharir2020Cost,Li2023Making} question the efficiency of traditional hardware, which lags behind that of biological systems by orders of magnitude \cite{Markovic2020Physics}. The efficiency gap is caused, i.a., by the hardware architectures internally working with digital signals, which are inconvenient for bio-inspired asynchronous operation, large data throughputs or analogue computations.
Therefore, a dedicated hardware will be used for neuro-inspired computing in the future. This will involve transferring certain high-level functions of traditional chips 
to physical systems and devices using their intrinsic characteristics such as temporal dynamics, mutual interactions, multi-state stability or memory \cite{Christensen20222022, Frenkel2023Bottom}. 
Various memristive \cite{Li2018Review}, optical \cite{Sui2020Review} or magnetic systems \cite{Kurenkov2020Neuromorphic} have been proposed to implement unconventional computing schemes such as reservoir computing or functions of synapses, neurons and whole networks.  
As for magnetic systems, this includes e. g. magnetic tunnel junctions  \cite{Leroux2021Hardware, Borders2019Integer, Krizakova2022Spin} and 
spintronic oscillators \cite{Torrejon2017Neuromorphic,Zahedinejad2020Two, Houshang2022Phase}, 
complex magnetic textures as skyrmions \cite{Raab2022Browniana, Zazvorka2019, Khodzhaev2023Analysis}
or chiral magnets and domain walls \cite{Bechler2023Helitronics, Hu2023Magnetic}, 
magnetic heterostructures \cite{KurenkovArtificial}, 
compensated magnets \cite{Bradley2023Artificial, Liu2022Compensated}, 
magnons \cite{Koerber2023Pattern} 
or magnetic metamaterials such as artifical spin ices \cite{Gartside2022Reconfigurable, Stenning2024Neuromorphic,Arava2019Engineering} 
or nanorings \cite{Vidamour2023Reconfigurable}.
Applications of these spin-based systems range from using AFM/FM spin-orbit devices used as weights \cite{Borders2016Analogue} for associative memory operation, over magnetic tunnel junctions employed as synapses and neurons for classification tasks \cite{Ross2023Multilayer, Romera2018Vowel} to implementing reservoir computing schemes with magnons for pattern recognition \cite{Koerber2023Pattern}.

Here, we explore the potential of an antiferromagnetic metal, CuMnAs, for unconventional computing applications.
   In previous works, magnetic properties  \cite{Wang2020Spin, zubac2021hysteretic, Janda2020Magneto} of CuMnAs films have been investigated including possible means of affecting the material's magnetic state \cite{Wadley2016Electrical, Olejnik2018Terahertz, Wadley2018Current, Amin2023Antiferromagnetic}. %
It was found, that the sample can be brought to a metastable high-resistive state by transient heating to the vicinity of Neél temperature ($\sim 480$~K) using an $\mu$s- to ns-scale electrical or fs-scale optical pulse \cite{Kaspar2020Quenching}. This effect, called quench switching, can be characterized by room-temperature dynamics at milliseconds and seconds.
Magnetic textures in CuMnAs and their control have been studied by microscopy methods \cite{Reimers2022Defect, Janda2020Magneto, Amin2023Electrical} and the atomically sharp 180$^{\circ}$ domain walls were identified as a possible microscopical origin of the high resistance signals \cite{Krizek2022Atomically}.
Works employing laser techniques have explored heat dissipation in CuMnAs layers at ultrashort timescales depending on their thickness \cite{Surynek2020Investigation} and
showed how to separate the transient heat contribution from the non-linear quench switching
and presented a phenomenological heat dissipation model \cite{surynek2024picosecond}.

In the present manuscript, we connect the short-term ultrafast sub-nanosecond heat dissipation phenomena, previously studied 
by advanced optical methods, with the quench switching, representing the long-term high-resistance memory to showcase neuromorphic information processing tasks.
First, in Fig.\ref{fig1}, we present the phenomenology of the quench switching,
combining a pair of excitation femtosecond laser pulses and electrical readout down to nanosecond time resolution to elucidate physical principles and neuromorphic functions provided by CuMnAs.
Second, in Fig.\ref{fig2}, we process a grayscale image by temporal latency coding,
encoding information from pixels into sub-ns delays between femtosecond laser pulses.
Next, we employ unequal laser pulses for pattern recognition and propose applying the same principle for convolutional edge detection.
Finally, we discuss the scalability needed for energy-efficient information processing and compare the antiferromagnetic CuMnAs with relevant magnetic and photonic systems used for solving similar neuromorphic tasks.

\section{Results} \label{sec:02_res}
 
\begin{figure}[ht]
 \centering
 \includegraphics[width=0.95\textwidth]{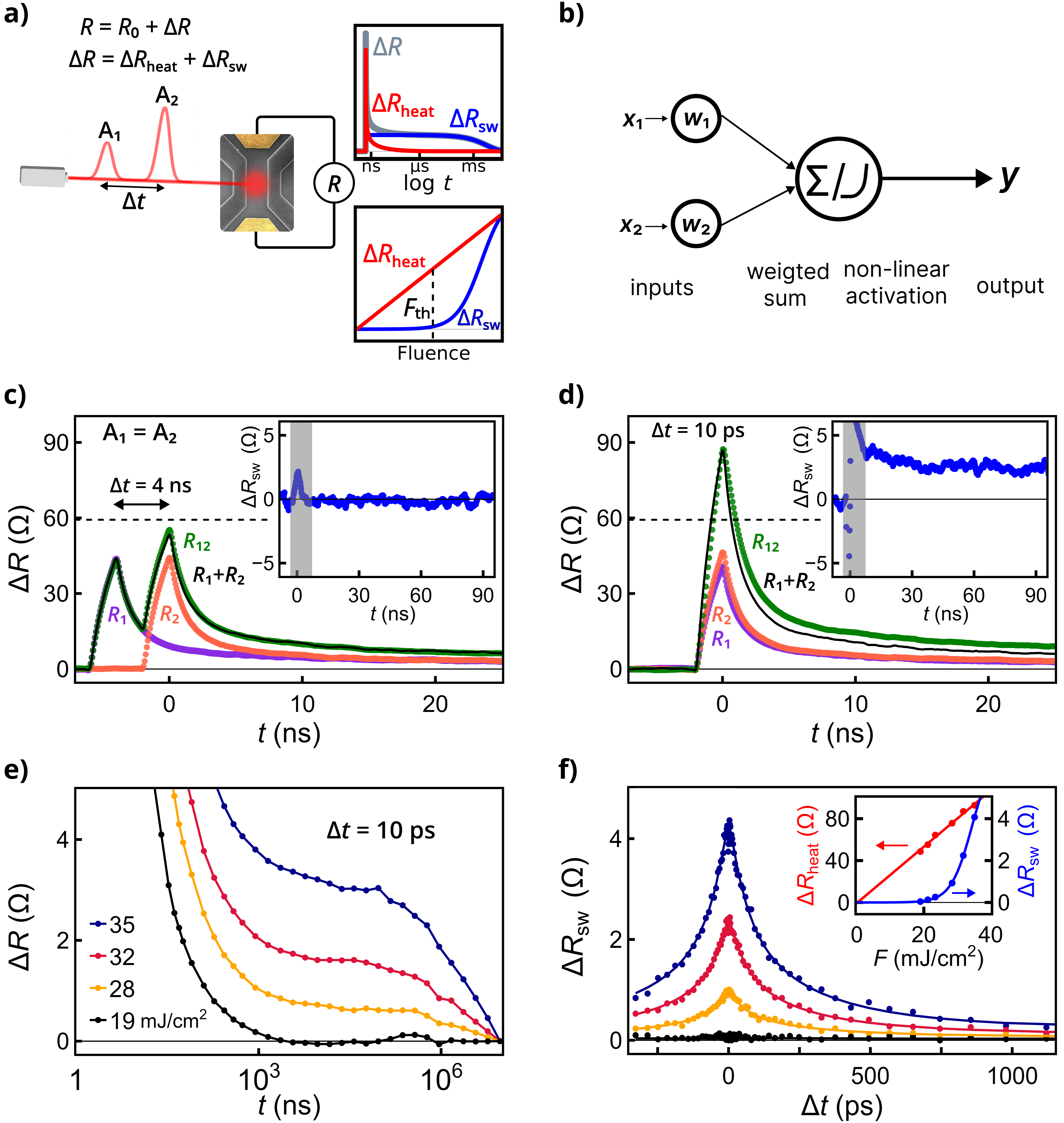}
 \caption{
 \footnotesize
 \textbf{Neuronal and synaptic functionalities realized in an antiferromagnet 
by optical switching using two laser pulses.}
 \textbf{(a)}~Design of the experiment: A 10-$\mu$m wide CuMnAs bar device prepared from a 20-nm epilayer is excited by two 150-fs input laser pulses of variable fluences delayed by~$\Delta t$. The full width at half maximum of the incident Gaussian beam is 20~$\mu$m. 
 A high-performance oscilloscope with $\sim$ns resolution was used to obtain the resulting longitudinal resistance $R$. The resistance $R$ consists of resistance offset $R_0$ and time-varying part $\Delta R_{\mathrm{sw}}$, comprising switching-signal ($\Delta R_{\mathrm{sw}}$) and heat ($\Delta R_{\mathrm{heat}}$) contributions (upper inset panel). $\Delta R_{\mathrm{heat}}$ and $\Delta R_{\mathrm{sw}}$ are fluence dependent (lower inset panel) with a characteristic onset in $\Delta R_{\mathrm{sw}}$ above switching fluence threshold $F_{\mathrm{th}}$.
%
\textbf{(b)} An artificial neuron taking two inputs. The neuron integrates weighted inputs $x_1$ and $x_2$ and produces an output once the weighted sum surpasses the threshold of an activation function. In the case of a dynamic neuron, integration is done over time and with a leakage.
 \textbf{(c)}~Resistance evolution at ns-timescale after the impact of two identical pulses delayed by 4~ns (green) and the same pulses applied individually (orange, purple). The black line is the ordinary sum of the orange and purple traces. Inset shows a resistance signal obtained by subtracting the sum from the experimental curve for both pulses.
 The grey area marks the region where the signal $\Delta R_{\mathrm{sw}}$ is ambiguous due to the uncertainty of the measurement method. The total fluence of the two pulses was 34.3~mJ/cm$^2$.
           \textbf{(d)}  Same as in (c) for the delay $\Delta t=10$~ps. The non-zero signal $\Delta R_{\mathrm{sw}}$ (inset) appears after the resistance 
           reaches the threshold for switching (dashed line). 
           \textbf{(e)}~Resistance evolution after the impact of two identical pulses delayed by $\Delta t = 10$~ps measured over the temporal range of 10~ms. A detail of the resistance part with prevailing switching contribution is shown for several cumulative fluences.
           \textbf{(f)}~Switching signal $\Delta R_{\mathrm{sw}}$ as a function of $\Delta t$ for a pair of pulses and same fluences as in~(e). Inset shows the experimental dependence of $\Delta R_{\mathrm{heat}}$ and $\Delta R_{\mathrm{sw}}$
           on the total fluence. The solid line is a fit of $\Delta R_{\mathrm{heat}}$ to a linear function, respectively a fit of $\Delta R_{\mathrm{sw}}$ to a sigmoid activation function. 
} 
\normalsize
 \label{fig1}
\end{figure}

In Fig.~\ref{fig1}, we present neuronal and synaptic functions in an antiferromagnet, employing switching by optical stimuli and electrical readout.
Experimentally, we excite the CuMnAs  
device prepared from a 20-nm-thick epilayer with two time-delayed 150-fs laser pulses, causing transient heating of the sample and recording its resistance with a high-frequency oscilloscope (Fig.~\ref{fig1}a). By varying the input parameters of the experiment - pulse fluences and their mutual delay - we control the ultrafast temperature evolution at nanosecond and sub-nanosecond timescales and thus determine the strength of the quench switching effect. This is reflected in the resulting resistance, which comprises time-varying heat contribution linear in fluence, $\Delta R_{\mathrm{heat}}$, and the non-linear quench switching contribution, $\Delta R_{\mathrm{sw}}$, characterized by onset above the threshold fluence, $F_{\mathrm{th}}$, and millisecond dynamics (insets of Fig.~\ref{fig1}a).
The arrangement corresponds to an artificial neuron, which integrates two weighted inputs in time and produces an output above a threshold (Fig.~\ref{fig1}b, see also Supplementary Fig.~S1  
for functionalities provided by the CuMnAs device). 
In Figs.~\ref{fig1}c,~d we experimentally explore the thresholding at nanosecond times by changing the pulse delay. 
The resistance data are shown for identical laser pulses delayed by 4~ns and 10~ps, 
with the fluences $A_1 = A_2 = 17.1$~mJ/cm$^2$.
The fluence value is selected to ensure that none of the pulses reaches the switching threshold individually ($A_i < F_{\mathrm{th}} \approx$ 22.2~mJ/cm$^2$), while the sum of pulse fluences exceeds the switching threshold ($A_1 + A_2 > F_{\mathrm{th}}$).
Due to the ultrafast heat dissipation on the order of $\sim 100$~ps, most of the heat imposed by the first pulse is already removed from the sample when the second pulse impacts 4~ns later. Consequently, the sample is not heated above the critical threshold value, and no switching is observed (Fig.~\ref{fig1}c).
As the delay between pulses is reduced close to the characteristic heat dissipation time or below, the heat contributions from the two pulses add up and accumulated heat surpasses the critical threshold value, resulting in a noticeable switching signal $\Delta R_{\mathrm{sw}}$ (Fig.~\ref{fig1}d for $\Delta t=10$~ps). 
The ability of heat accumulation combined with the exponential decay in time due to dissipation is analogous to the leaky integration in the leaky-integrate-and-fire neuron model. 
To distinguish the signal $\Delta R_{\mathrm{sw}}$ (insets of Figs.~\ref{fig1}c, d) from $\Delta R_{\mathrm{heat}}$ at the nanosecond times, where the two contributions overlap, we estimated the transient heat contribution by the sum of traces for individual sub-threshold pulses $(R_{1}+R_{2})$, which we subtracted from the trace for both pulses applied simultaneously ($R_{12}$), i.~e. $\Delta R_{\mathrm{sw}}= R_{12} - \Delta R_{\mathrm{heat}} \approx R_{12} - (R_{1}+R_{2})$.  
At times after impact longer than $\approx \mu$s, the heat has already been fully
dissipated from the sample to the substrate and only the switching signal $\Delta R_{\mathrm{sw}}$ adds up to $R_0$. Hence, the amplitude of the switching signal can be evaluated directly from the traces measured over a 10-ms window by fitting to the stretched exponential function $\sim \exp \left( (-t/\tau)^\beta \right)$ for $t \gtrsim \mu$s with $\tau = 4$~ms. 
In this way, information about sub-nanosecond pulse delays can be retrieved at milliseconds,
providing temporal scalability across several orders of magnitude without the necessity of implementing ultrafast readout. 
This approach can be conveniently employed for representing and processing data, as we will show further.

Beside changing delay between pulses, we can also control the heat accumulation and switching by adjusting their fluences.
The fluence-dependent resistance traces in the 10-ms window range for a constant pulse delay $\Delta t = 10$~ps are presented in Fig.~\ref{fig1}e. Whereas the resistance trace for 19~mJ/cm$^2$ is dominated by heat contribution, approaches $R_0$ after the heat fades away and shows no switching signal, the other traces exhibit the switching manifested by increased resistance at microsecond and millisecond times. 
Fig.~\ref{fig1}f then summarizes previous figures by displaying the switching signal $\Delta R_{\mathrm{sw}}$ for various fluences as a function of the pulse delay $\Delta t$.
It has been shown that the observed dependence can be explained by the heat dissipation model comprising several contributions of various origins \cite{surynek2024picosecond}. The variation at $\sim 100$-ps timescales can be attributed to ultrafast heat dissipation mediated by phonons. Although the heat dissipation time constant typical for the 20-nm CuMnAs film is $\sim 500$~ps \cite{Surynek2020Investigation}, $\Delta R_{\mathrm{sw}}$ varies more rapidly than that ($\sim 200$~ps), further enhancing the temporal resolution provided by the switching signal.  The distinction is the consequence of the $\Delta R_{\mathrm{sw}}$ dependence on fluence, which is, in contrast with $\Delta R_{\mathrm{heat}}$, non-linear, as highlighted by the experimental data in the inset of Fig.~\ref{fig1}f. 

\begin{figure}[ht]
 \centering
 \includegraphics[width=0.92\textwidth]{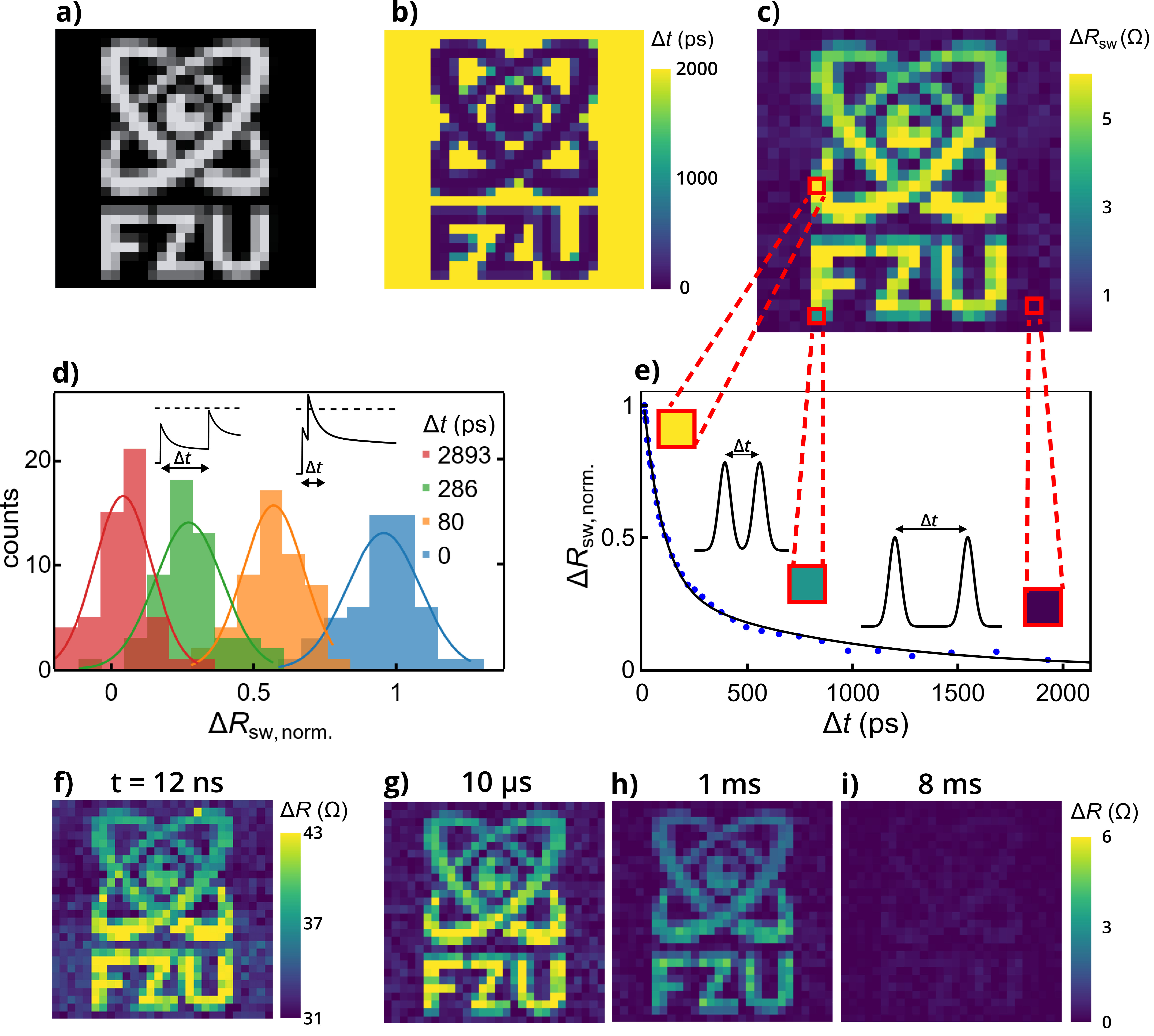}
 \caption{
 \textbf{Temporal latency coding of data from an image.}
           \textbf{(a)}~Original greyscale image with 256 colour levels.
           \textbf{(b)}~Input ns-scale delays. We used a monoexponential fit to the experimental $\Delta R_{\mathrm{sw}}$ vs. $\Delta t$ relation to determine the delays, keeping the resulting switching signal $\Delta R_{\mathrm{sw}}$ linear with the greyscale of the original image. The delays $\Delta t$ employed ranged from 10 to 2000~ps. 
           \textbf{(c)}~Switching signal measured at 10$\mu$s carrying information from the image and correspondence to the experimental $\Delta R_{\mathrm{sw}}$ vs. $\Delta t$ dependence~\textbf{(e)}. The experiment used two laser pulses of equal amplitude and total fluence 33 mJ/cm$^2$. 
           The switching signal was normalized to the value for the zero delay.
           \textbf{(d)}~Distributions of switching signal in repeated experiments for selected pulse delays $\Delta t$. The plot was created from the data in Fig.~\ref{fig1}f using the three curves showing a nonzero signal.  
           \textbf{(f)-(i)} Short- to long-term memory transfer and memory fading demonstrated by resistance readout at different times $t$ after the impact of the pulses. Short-term heat-dominated resistance information \textbf{(f)} is transferred into the long-term switching signal \textbf{(g)}, which fades out over a 10-ms period \textbf{(g-i)}, after which the memory device can be reused. 
 } 
 \label{fig2}
\end{figure}

Combining the ultrafast heat dynamics triggered by time-delayed femtosecond laser pulses and the antiferromagnetic memory opens up pathways for sub-nanosecond information processing, as presented in Fig.~\ref{fig2}. The technique used is called temporal latency coding:
We encode the information from a greyscale image (Fig.~\ref{fig2}a) into mutual time delays of two equal pulses (Fig.~\ref{fig2}b). Using the previously acquired experimental $\Delta R_{\mathrm{sw}}$ vs $\Delta t$ dependence (Fig.~\ref{fig2}e), the image of dimensions 28$\times$28 pixels is converted into delays in such a way that the image's 256 levels of grey are linearly mapped onto the range of $\Delta R_{\mathrm{sw}}$. The appropriate delay $\Delta t$ is then obtained from $\Delta R_{\mathrm{sw}}$ as the inverse function. This procedure led in our setting to delays from 10 to 2000~ps. 
Next, we perform the two-pulse switching experiment sequentially using these delays as inputs. From the measured resistance traces, we evaluate the map of switching signal $\Delta R_{\mathrm{sw}}$ as presented in Fig.~\ref{fig2}c. The resulting $\Delta R_{\mathrm{sw}}$ map reproduces the original image (Fig.~\ref{fig2}a) and captures the information from the sub-ns delay map (Fig.~\ref{fig2}b), but it has an opposite contrast, i.e., shorter delays lead to a larger signal. Contrary to the ultrafast pulse delays, the information in the switching signal can be retrieved at microseconds and milliseconds after the pulses' impact using a conventional electrical readout.

In Fig.~\ref{fig2}d, we study the reliability of the used encoding procedure by recognizing different pulse delays from the switching signal, $\Delta R_{\mathrm{sw}}$. We selected four different delays from 0 to $\sim 3000$~ps using the experimental data from Fig.~\ref{fig1}f for fluences showing a non-zero switching signal in the pulse overlap. The signal was always normalized to a value of maximal signal reached for short pulse delays and fitted to a Gaussian function. The underlying data distributions had standard deviations from 0.10 to 0.12 of normalized $\Delta R_{\mathrm{sw}}$,
illustrating the possible discernible states and the noise of the data that determine the maximum information transfer rate (see Shannon-Hartley theorem, e.g. in Ref.~\cite{Singh2022survey}).
 Figs.~\ref{fig2}f-i show the relation between the heat and switching signal contributions to the resistance and following signal fading by comparing the resistance traces at different laboratory times. Whereas short after the pulses' impact, the resistance variation $\Delta R$ is dominated by the heat contribution (see Fig.~\ref{fig2}f for $t=12$~ns), the remaining Figs.~\ref{fig2}g-i at millisecond times show solely the switching signal. 
 By comparing Fig.~\ref{fig2}f and Fig.~\ref{fig2}g, it is apparent that the information from the heat is translated into the switching signal. Additionally,  Figs.~\ref{fig2}g-i show the signal variation at $\mu$s and ms times. 
The illustrated signal decline and its vanishing, given by the memory's metastability and switching signal dynamics, are crucial for time-division multiplexing - the capability to reuse the device again after the memory's forgetting period without the necessity of resetting the system's magnetic state by field or thermal cycles. 
The switching signal fading memory and the information transfer across orders of magnitude in time due to the combination of the short-term heat-related memory and the long-term switching-signal memory are remarkable features demonstrating the system's information encoding capabilities. 

\begin{figure}[ht]
 \centering
 \includegraphics[width=\textwidth]{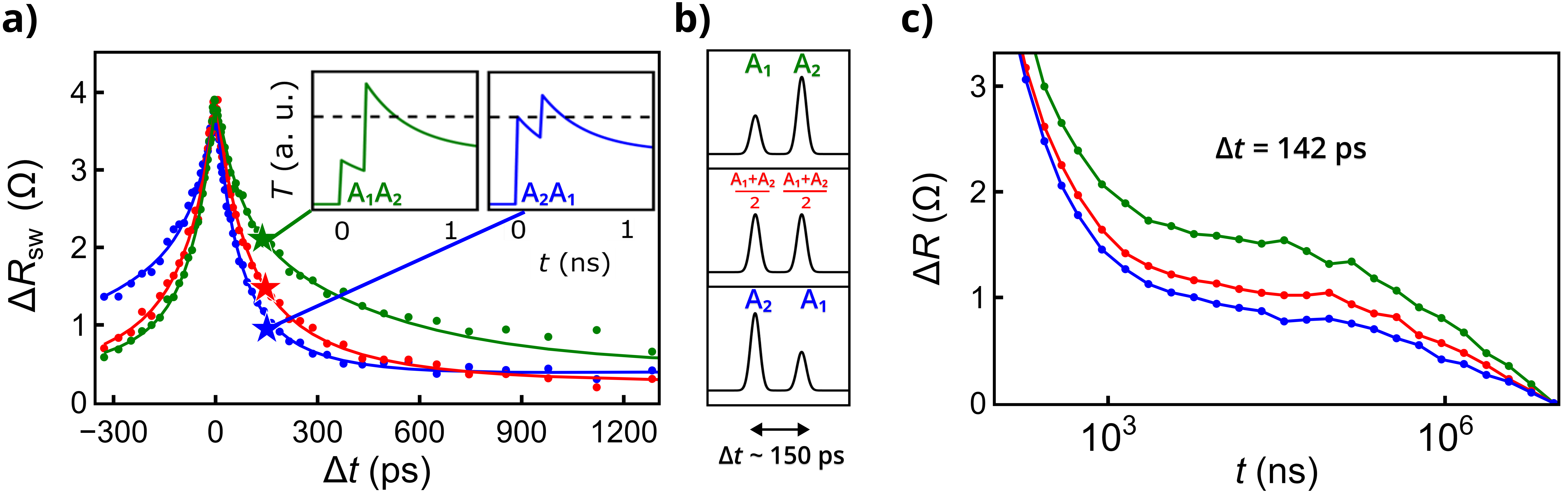}
 \caption{ \textbf{Recognizing temporal patterns of pulses.} 
 \textbf{(a)}~Switching with uneven excitation pulses with the same total energy but different ratios of individual pulse fluences can be used for recognizing pulse patterns at $\sim$~100-ps timescales. The inset shows the device's temperature at sub-ns times for sequences of uneven pulses of opposite order $A_1A_2$ vs. $A_2A_1$. Pulses of total commulative fluence $A_1 + A_2 \approx 33$~mJ/cm$^2$ 
 with ratios of pulse fluences $ \frac{A_1}{A_2} = \frac{2}{1}$, $\frac{1}{1}$, and $\frac{1}{2}$ were employed as illustrated in \textbf{(b)}. \textbf{(c)}~Average experimental resistance traces for the three pulse ratios and the delay $\Delta t = 142$~ps. 
 } 
 \label{fig3}
\end{figure}

\begin{figure}[ht]
 \centering
 \includegraphics[width=\textwidth]{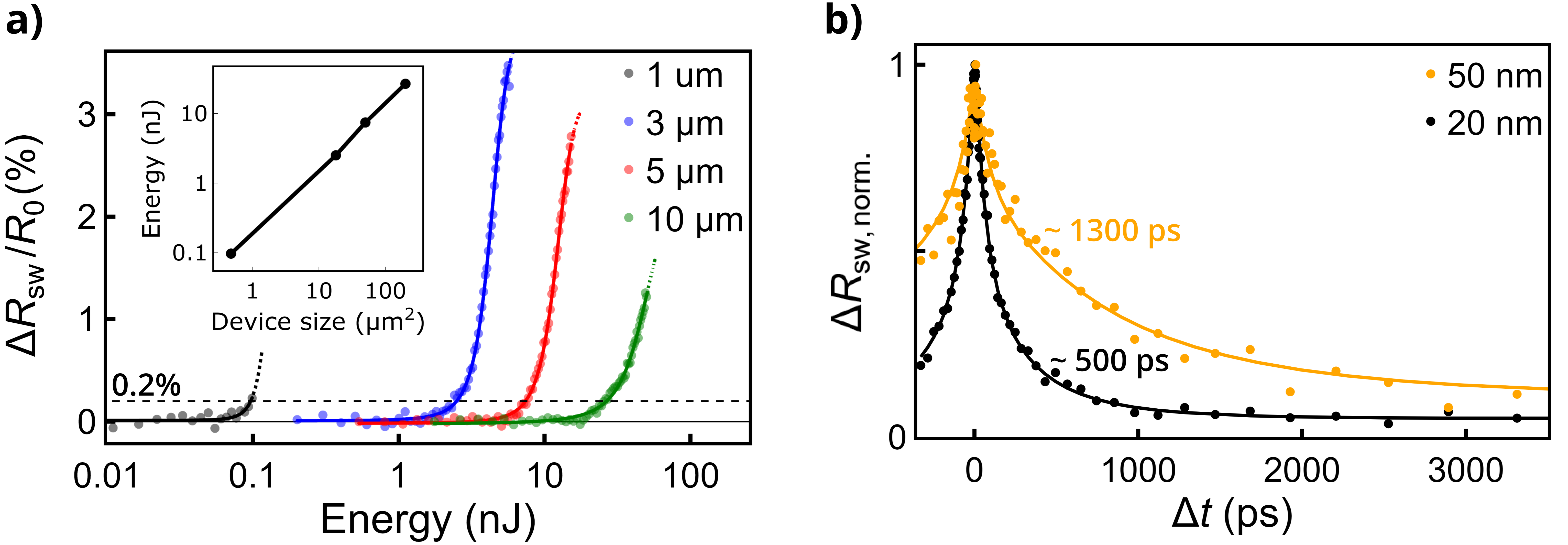}
 \caption{ \textbf{Scalability and energy efficiency.} 
 \textbf{(a)}~Scalability of the devices illustrated by single pulse switching for devices of different nominal widths $w$ prepared from the 20-nm CuMnAs film. For all device sizes, the switching signal dependence on energy adopts the same functional form of a sigmoid. Inset shows the energy of the pulse incident onto the samples' active regions needed to reach the relative switching signal of $\Delta R_{\mathrm{sw}}/R_0 = 0.2 \%$. 
 The full width at half maximum of the incident Gaussian laser beam was 20~$\mu$m.
 Nominal dimensions of the active region ($\sim w \times 2w$ $\mu$m$^2$) were corrected according to their actual size after preparation by wet etching in the case of the smallest device. For larger devices ($w \gtrsim 3$~$\mu$m), these corrections are on the order of $\sim 1 \%$ and thus negligible in the presented log-scale.
 \textbf{(b)}~Scaling of device thicknesses leads to different thermal heat dissipation rates as shown by normalized $\Delta R_{\mathrm{sw}}$ vs. $\Delta t$ dependence for the 50-nm and 20-nm thick devices with the dimensions of the active region 10 $\times$ 20 $\mu$m$^2$.
 } 
 \label{fig4}
\end{figure}

Next, in Fig.~\ref{fig3}, we show that our system can be used not only for the information encoding into delays but also for recognizing patterns of pulses at ultrashort times when distinct pulse amplitudes are employed. Fig.~\ref{fig3}a shows the sub-ns delay dependence of the switching signal $\Delta R_{\mathrm{sw}}$ for two pulses of different fluence ratios but the same sum of fluences. Here, the pulse fluences are again selected to lie below the switching threshold individually but exceed the threshold in total. 
In contrast with the reference equal-pulse case symmetric with respect to $\Delta t$ to $-\Delta t$ conversion (red curve), the switching signal for a given pulse delay is larger when the weaker pulse is followed
by a stronger (green curve and $\Delta t>0$, blue curve and $\Delta t<0$) and lower when the stronger pulse is followed by a weaker one (green curve and $\Delta t<0$, blue curve and $\Delta t>0$). 
The manifested behaviour can be understood considering the ultrafast heat dissipation in the material. 
The temperature evolution after the impact of the pulses can be approximated \cite{surynek2024picosecond} as
\begin{equation}
   \Delta T (t)  \approx A_1  \exp^{+}(-\frac{t}{\tau_\mathrm{ph}} ) + A_2  \exp^{+}(-\frac{t-\Delta t}{\tau_\mathrm{ph}} ),
    \label{eq0}
\end{equation}
where $A_1$ and $A_2$ are the pulse amplitudes, $\tau_\mathrm{ph} \sim 500$~ps is the heat dissipation time and $\exp^{+}(-x)=\exp(-x)\Theta(x)$ is the exponential function multiplied by the Heaviside step function $\Theta(x)$. Since the amount of the dissipated heat scales with the pulse amplitude,
the temperature generally reaches a different value at $t=\Delta t$ for the opposite order of the
pulses. In our setting, where the ratios of pulse amplitudes $A_1$ and $A_2$ are $\frac{1}{2}, \frac{2}{1}$ or $\frac{1}{1}$ and their sum is $A_1 + A_2 \approx 1.48\, F_{\mathrm{th}}$ , 
it leads to a different maximal temperature reached after the impact of the second pulse (see the inset of Fig.~\ref{fig3}a).  
This difference is translated into different values of the switching signal being recorded at ms-times since its amplitude is a function of the maximal temperature, with a non-linear onset at $F_{\mathrm{th}}$. Additional discussion of the heat dissipation and the switching behaviour is provided in Section~2 
of the Supplementary material. 
In Fig.~\ref{fig3}c, we present the respective experimental resistance traces for the three fluence ratios recorded in the 10-ms window and the selected pulse delay $\Delta t = 142$~ps.
The curves further demonstrate switching signals dependence on the distribution of the total fluence into fluences of individual pulses.
It shows that the switching signal is larger for a sequence $A_1A_2$ of a small pulse followed by a larger one than for the opposite pulse order.
The accuracy of recognizing pulse patterns consisting of two pulses with amplitudes $A_1$ and $A_2$, can be as high as $\sim 98\%$ as we corroborate in supplementary Fig.~S3 
using all four combinations of $A_1$ and $A_2$ and  $\Delta t= 150$~ps.
The presented recognition of patterns using two pulses can be further used as a basis for more subtle processing, e.g., it can be extended into a case of multiple pulses representing additional inputs, or it can be used to perform more complex operations. In supplementary Fig.~S4,
we propose the application of pulse-order recognition for convolutional edge detection using a binary image as the input data. 
Finally, we discuss the prospects of the antiferromagnetic CuMnAs material for designing novel computing architectures by studying devices' lateral and vertical scalability and energy efficiency. Fig.~\ref{fig4}a shows the relative single-pulse switching signal $\Delta R_{\mathrm{sw}}/R_0$ plotted against the energy of the laser pulse incident onto the samples' active region for devices of various sizes. 
The devices were prepared from the 20-nm-thick CuMnAs film, and the dimensions of their active region were approximately $w \times 2w \, \mu$m$^2$, where $w$ is the device width. The full width at half maximum of the incident laser beam was 20~$\mu$m, always safely covering the entire active region.
As the dimensions are scaled from the 10-$\mu$m-wide bar down to sub-1-$\mu$m$^2$ device, the amount of energy impacting the device needed to reach the switching signal $\Delta R_{\mathrm{sw}}/R_0 = 0.2\%$ 
drops from $\sim 10$~nJ to sub-$0.1$-nJ values. Despite scaling, the multilevel character of the switching and temporal relaxation of the signal is preserved, indicating that the magnetic textures in CuMnAs responsible for the quench switching signals are much smaller than the size of the smallest prepared device, i.e. $0.5 \mu$m$^2$.
Indeed, the previous microscopy study based on DPC-STEM and micromagnetic XMLD-PEEM imaging \cite{Krizek2022Atomically} revealed a potential source of the quench switching signal in spin textures and sharp domain walls at the atomic scale, well below the detection limit of the XMLD-PEEM method ($\sim 10$~nm). Extrapolating the nearly linear dependence of threshold energy on the device size (inset of Fig.~\ref{fig4}a) down to $\sim 10 \times 10 = 100$~nm$^2$ dimensions
leads to an energy estimate of $\sim 10$~fJ for such a minuscule, yet still conceivable device.
The value of $\sim 10$~fJ  
needed for switching lies beyond the current limits of the energy cost per floating point operation in modern CMOS microprocessors ($\sim 150$~fJ, \cite{Ho2023Limits}),
is commensurate with the most energy-efficient magnetic memories used for neuromorphic computing
( $\lesssim 100$~fJ \cite{Wijshoff2024Enhanced,Rao2023Spin,Grollier2020Neuromorphic}) 
or state-of-the-art organic artificial synapses \cite{Lee2021Organic}, and comparable with characteristic energy per synaptic event in biological systems ($\sim 1-100$~fJ, \cite{Burgt2017non}).
Further improvement of energy efficiency and heat dissipation properties of devices 
can be achieved by scaling the vertical dimensions of layers and reducing their thickness. %
Fig.~\ref{fig4}b compares the delay dependence of $\Delta R_{\mathrm{sw}}$, given by the heat dissipation, in the 50-nm and 20-nm CuMnAs film. Whereas in the thicker sample, the thermal relaxation takes more than 1~ns, the heat in the 20-nm sample is dissipated within $\sim 500$~ps. The previous report \cite{Surynek2020Investigation} consistently demonstrated heat dissipation times approximately proportional with the layer thickness, with values $\lesssim 100$~ps for the thinnest, less than 10~nm thick, layers. 
Moreover, decreased film thickness improves heat transfer efficiency not only through the reduction of heat dissipation times but also by limiting the relative amount of residual heat accumulated in the sample \cite{surynek2024picosecond}, enabling possible denser integration of devices and increasing their endurance by mitigating thermal degradation.
%


\section{Discussion} \label{sec:03_dis}
In this article, we explored synaptic and neuronal functionalities of the antiferromagnetic CuMnAs device for scalable and energy-efficient information processing using the electrical readout and a pair of femtosecond laser pulses as the input stimulus. 
We demonstrated the encoding of a grayscale image by temporal latency coding, transferring information from sub-ns delays to switching signal at milliseconds (Fig.~\ref{fig2}).
We also showed how to use the combination of ultrafast heat dynamics and quench switching of an antiferromagnet for recognizing pulse patterns at $\sim 100$~ps timescales (Fig.~\ref{fig3}) and proposed utilization in convolutional edge detection (Supplementary Fig.~S4).
Let us now review several physical systems and devices exploiting related techniques and principles which were used to address similar tasks and compare them with the results of this article. 

We start with the observation from Fig.~\ref{fig4} that both the 10$\mu$m-wide CuMnAs device and a much smaller, sub-1-$\mu$m$^2$ device exhibit the same multilevel field-insensitive switching characteristics due to expected atomic-level sizes of the underlying magnetic textures, highlighting devices' ultimate scalability.
Previous works employing compensated magnets for neuromorphic applications relied on features of micrometer-sized domains with implications for the functionalities. In Ref.~\cite{KurenkovArtificial} studying neuronal functions realized by spin-orbit torque switching in the antiferromagnet/ferromagnet heterostructure, the multistability was attributed to an intermediate state due to local exchange bias variations at $\sim 200$~nm scales leading to a transition from multistable switching for the 5-$\mu$m device to probabilistic binary switching for the device smaller than 1~$\mu$m. Moreover, the magnetic field was required for reinitiation of the device's magnetic state and its repeated use.
The article \cite{Liu2022Compensated} presented synaptic behaviour for ultrafast neuromorphic computing in the compensated ferrimagnet switched under magnetic field by spin-orbit torques.
Gradual changes of synaptic weights were realized by electrical switching with 10-ns pulses and $\sim 30$~pulses were needed for the full magnetization reversal.
Both works used electrical pulses with $\geq 1$~ns duration, which limits the temporal resolution for switching by the coincidence of two pulses. On the contrary, the method presented in this article provided a resolution at $\sim 100$-ps times permitted by the ultrafast heat dissipation in CuMnAs layers (Figs.~\ref{fig1}f and \ref{fig4}b).

As for applications of physical systems for solving tasks, pattern recognition based on non-linear scattering of magnons was demonstrated in a magnonic Ni-Fe reservoir excited in the high-power regime at GHz frequencies \cite{Koerber2023Pattern}. Two microwave pulses of distinct frequencies with the 20-ns envelope were used as the excitation, and Brillouin light spectroscopy was used as the readout. The maximal difference between the spectra corresponding to the opposite pulse order, connected to the highest pattern recognition efficiency, was recorded in the case of partially overlapped pulses with $\Delta t \sim$~10ns.
Furthermore, in Ref.~\cite{Robertson2020Ultrafast} authors presented ultrafast photonic neuromorphic image processing using a single neuron represented by a commercially available vertical cavity surface emitting laser (VCSEL). Similarly to our approach, the authors take advantage of the temporal integration of laser pulses, causing the production of neuron output above a threshold, which is, in their case, reached only when all the input pulses have the same negative polarity. The method was applied for convolutional edge detection with $2\times2$ kernels using 3-ns delays between input laser pulses, consuming 1.2~pJ per operation.
In this work, analogous functionalities such as in Refs.~\cite{Koerber2023Pattern, Robertson2020Ultrafast} has been realized by the new method based on quench switching in an antiferromagnet, with competitive energy efficiency and shorter temporal pulse separation ($\sim 100$~ps). Moreover, both excitation and readout can be performed optically as well as electrically in our system (Ref.~\cite{Kaspar2020Quenching} and Supplementary Fig.~S6), further highlighting the versatility by enabling all-optical operation with free-space connectivity on the one hand, and more accessible all-electrical operation on the other. 
The recent work \cite{Olejnik2024} also confirmed the same switching effect as in CuMnAs in a different, higher Néel temperature antiferromagnet, allowing for enhancing device properties for neuromorphic computing through material optimization.


\section{Conclusion}
In summary, we studied a novel concept for neuromorphic information processing using an antiferromagnet, combining the short-term ultrafast heat dynamics and the long-term quench switching. The two phenomena of distinct dynamics can be used for temporal latency encoding, inherent information representation in spiking neural networks.
Recognition of patterns, achieved at $\sim 100$~ps by employing pulses of different amplitudes, is then an essential operation for various data processing tasks,  with applications in computer vision or signal analysis. Experimental realization of these functions at ultrashort times paves the way for efficient time-multiplexed in-materia computing using compensated magnets.

\section{Experimental Section}

\subsection{Sample preparation and device fabrication}
Thin films of tetragonal CuMnAs with 20 and 50 nm thickness were grown by molecular epitaxy on GaP(001) substrate. Simple two-terminal bars used for experiments were prepared by electron beam lithography and wet etching. 
Then, we used the evaporation technique to equip the bars with golden contacting pads, assuring similar resistance of devices from the same batch, independently on exact contact positioning.
 Further details about the thin film characterization can be found in 
Ref.~\cite{Krizek2020Molecular}.

\subsection{Optical measurements}

The optical experiments were performed using a femtosecond laser system Pharos by Light Conversion. The laser pulse duration was 150~fs and the central wavelength was 1030~nm.
A sketch of the experimental setup can be found in the Supplementary material (Fig.~S4).
Details about the experimental setup can be found in Ref.~\cite{surynek2024picosecond}.

\subsection{Electrical measurements}
Time-resolved electrical measurements were performed using a Rohde \& Schwarz RTP064 high-performance oscilloscope with 6~GHz bandwidth for readout and Rigol DG1000Z as a voltage source. We recorded the voltage $U$ of the input resistance of the oscilloscope connected in series with the sample and calculated the sample's resistance.  

We typically collected 2.5M points over a 10~$\mu$s or a 10~ms window during one acquisition, which was started by receiving a timing reference signal from the laser.
To reduce the noise at high frequencies without losing the time resolution, we acquired the data in a stroboscopic regime with a 100-Hz laser duty cycle and averaged out curves from $\sim$100 acquisitions. Schematics of the experimental setup can be found in the Supplementary material (Fig. S4).

\medskip
\subsection*{Supporting Information} \par 
Supporting Information is available from the Wiley Online Library or from the author.

\medskip
\subsection*{Acknowledgements} \par 
T.J. acknowledges Ministry of Education of the Czech Republic Grant No. CZ.02.01.01/00/22008/0004594 and LM2023051 and ERC Advanced Grant No. 101095925.
J.Z., K.O., A.F., F.K, and Z.K. acknowledge support from Czech Science Foundation Grant No. 21-28876J.




\clearpage
\bibliographystyle{unsrt} 
\bibliography{mybibliography_spintronics} 

\end{document}


\beginsupplement

\title{\textbf{Neuromorphic information processing using ultrafast heat dynamics and quench switching of an antiferromagnet: Supplementary material}}  
\date{}
\author[1,2]{J. Zubáč\footnote{Email: zubac@fzu.cz}\footnote{These authors contributed equally.}}
\author[2]{M. Surýnek$^{\dagger}$}
\author[1]{K. Olejník}
\author[1,2]{A. Farkaš}
\author[1]{F. Krizek}
\author[2]{L. Nádvorník}
\author[2]{P. Kubaščík}
\author[1,2]{Z. Kašpar}
\author[2]{F. Trojánek}
\author[3]{R. P. Campion}
\author[1]{V. Novák}
\author[2]{P. Němec}
\author[1,3]{T. Jungwirth}
\affil[1]{Institute of Physics of the Czech Academy of Sciences, Prague, Czech Republic}
\affil[2]{Faculty of Mathematics and Physics, Charles University, Prague, Czech Republic}
\affil[3]{School of Physics and Astronomy, University of Nottingham, United Kingdom}

\maketitle  

\vspace{-0.66cm}
\section{Schematics of a neuron and neuromorphic functionalities}
\label{supp_sec1:schematics}

\begin{figure}[h!]
 \centering
 \includegraphics[width=0.93\textwidth]{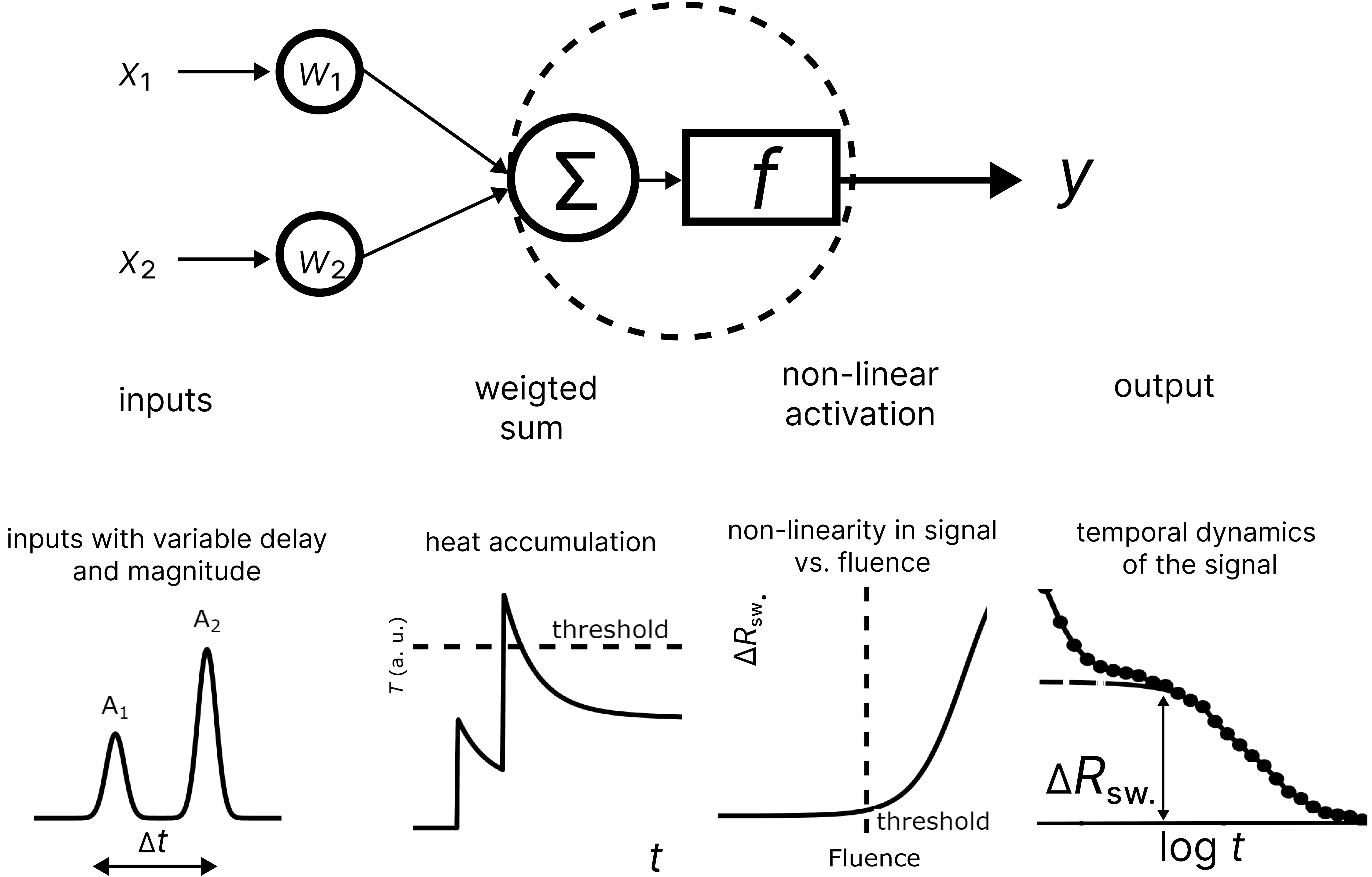}
 \caption{ \textbf{Schematics of a neuron and neuromorphic functionalities.}
           An artificial neuron takes two inputs, performs their weighted summation and passes the result through a non-linear activation function. Analogous functionalities in the antiferromagnetic memory device: Two excitation laser pulses with variable delays and magnitudes heat the sample. Once the temperature threshold is reached, switching signal $\Delta R_{\mathrm{sw.}}$ is observed. Twofold dynamics - short-term heat dynamics and long-term switching dynamics characterize the behaviour of the sample.  
} 
 \label{figS1}
\end{figure}

\vfill  

\section{Heat dissipation model for two unequal pulses}
\label{supp_sec2:heat_dissip}
To explain the observed behaviour from Fig.~3 
used for the recognizing temporal patterns of pulses, we extend the previously presented heat dissipation model using two pulses of the same magnitude from Ref.~\cite{surynek2024picosecond} into a case when the excitation laser pulses are unequal.
The dissipation of laser-induced heat into a metallic sample at picosecond times is usually described by a phenomenological two-temperature model considering electron and phonon subsystems of a sample with different temperatures. Following the absorption of laser pulse energy by the electron subsystem and subsequent electron thermalization, the phonon and electron subsystem's temperatures equalize with a characteristic time constant  $\sim 2.4$~ps \cite{Surynek2020Investigation}. At longer times, the thermal relaxation mediated by phonons with $\sim 100$-ps dynamics
dominates the heat dissipation process, followed by the heat diffusion from the 
substrate to other thermally connected parts of the experimental setup (sample holder, etc.)
with an observed characteristic time constant $\tau_\mathrm{subst} \sim 10$~ns.
In the case of a single pulse hitting the sample at the time $t=0$, this process can be described by a function
\begin{equation}
    f (t) = (A - B) \exp(-\frac{t}{\tau_\mathrm{ph}}) + B \exp(-\frac{t}{\tau_\mathrm{subst}}),
    \label{eq1}
\end{equation}
where $\tau_\mathrm{ph}$ and $\tau_\mathrm{subst}$ are the time constants, $A$ is the amplitude of the laser pulse, $A-B$ represents a portion of heat dissipated with the dynamics given by the $\tau_{\mathrm{ph}}$ and $B$ is the heat contributing to overall heating of the substrate.
Assuming the maximal temperature change is approximately proportional to the absorbed laser fluence, the function $f (t)$ can be interpreted as the sample's temperature change.
Considering two laser pulses, the first impacting the sample at time $t=0$ and the second at $t=\Delta t$, the temperature's time evolution can be roughly approximated as

\begin{equation}
   \Delta T_{12} (t) = f_1 (t) + f_2(t-\Delta t) \approx A_1  \exp^{+}(-\frac{t}{\tau_\mathrm{ph}} ) + A_2  \exp^{+}(-\frac{t-\Delta t}{\tau_\mathrm{ph}} ),
   \label{eq2}
\end{equation}
where $\exp^{+}(-x)=\exp(-x)\Theta(x)$ is the exponential function multiplied by the Heaviside step function $\Theta(x)$.

\begin{figure}[ht]
\centering
\includegraphics[width=0.92\textwidth]{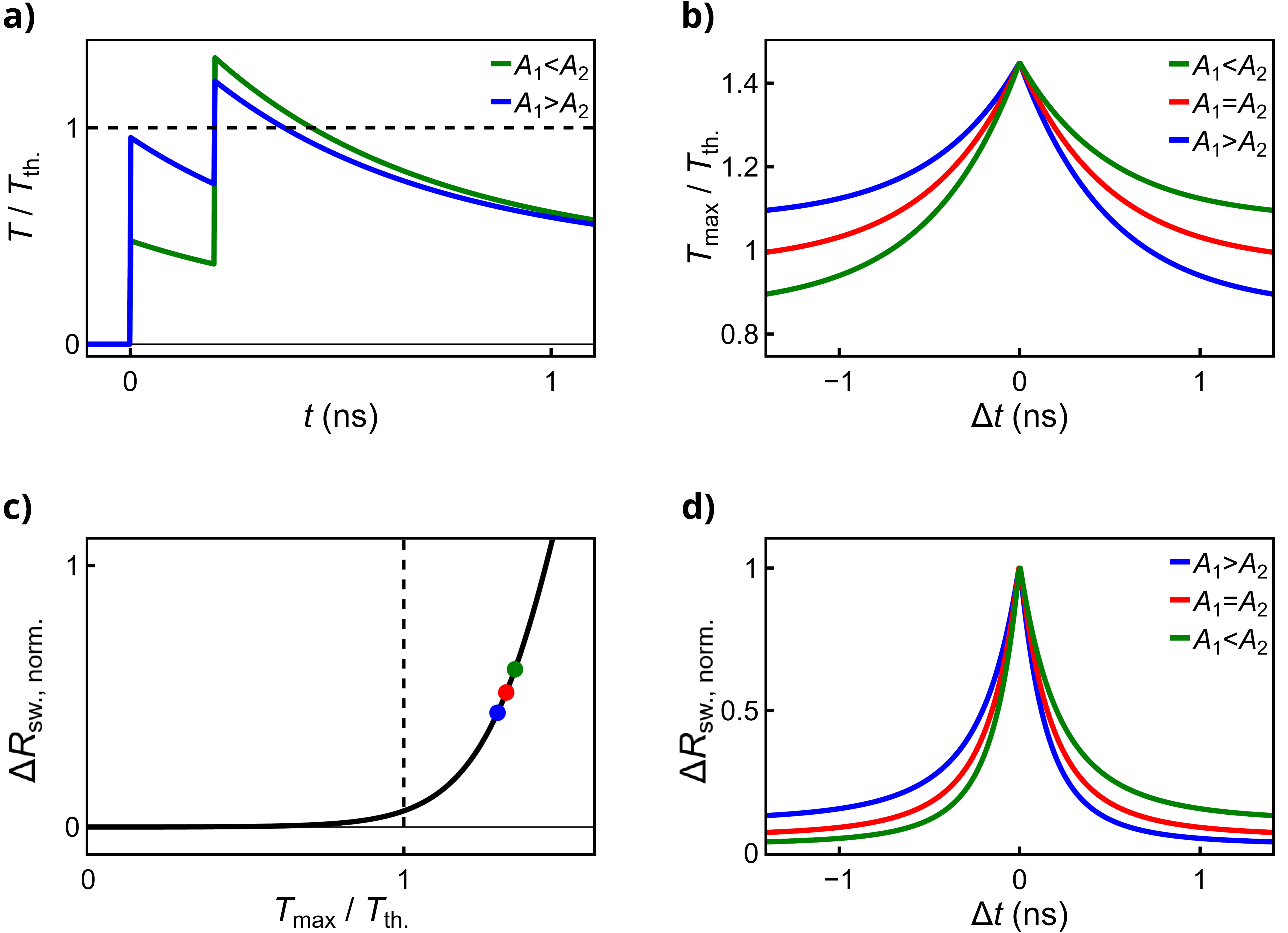}
\caption{ \textbf{Calculated sample temperature after the impact of two laser pulses of various fluence amplitude ratios and the corresponding switching signal.} Dissipation rates $\tau_{\mathrm{ph}} = 0.5$~ns and $\tau_{\mathrm{subst}} = 10$~ns, and amplitudes $A_1 + A_2 = 1.45\, F_{\mathrm{th}}$ were used. The ratio of pulse amplitudes is $\frac{A_1}{A_2}=\frac{1}{2}$ for $A_1 < A_2$ 
and $\frac{A_1}{A_2}=\frac{2}{1}$ for $A_1 > A_2$. Parameters of the sigmoid were obtained from the fit to experimental the data (inset of Fig.~1f). 
\textbf{(a)} Temporal dependence of the sample temperature after the impact of two pulses with the same fluences but opposite order and the delay $\Delta t =150$~ps .  \textbf{(b)} Delay dependence of the sample temperature for the three pulse ratios. \textbf{(c)} Nonlinear sigmoidal dependence of the switching signal on the maximal temperature. The signal value was normalized to the value reached in the pulse overlap. The coloured dots represent the signal values for the three amplitude ratios depicted in \textbf{(b)} and the pulse delay $\Delta t =150$~ps. \textbf{(d)} The resulting switching signal as a function of pulse delay.
}
\label{figS5}
\end{figure}

In Fig.~\ref{figS5}, we simulate the sample temperature using equations Eq.~\ref{eq1} and Eq.~\ref{eq2} during the double pulse experiment and illustrate the resulting switching signal. 
Fig.~\ref{figS5}a shows the temperature change for two pulses of the same pulse magnitude but
opposite order as a function of laboratory time, $t$. 
Since the amplitudes $A_1$ and $A_2$ in Eq.~\ref{eq2} are weighted by exponential dissipation factors shifted in time, the heating by two pluses in times $t \lesssim \tau_\mathrm{th}$ 
does not correspond to a combined effect as if the pulse fluences were summed, 
but leads to a higher maximal temperature when the smaller pulse is followed by a larger than in the opposite case.
%
%
%
%
%
In Fig.~\ref{figS5}b, we plot the corresponding dependence of maximum reached temperature on the pulse delay, $\Delta t$, for different fluence ratios but the same sum of pulse fluences. The parameters
used were the same as in  Fig.~\ref{figS5}a; we added the reference curve for $A_1=A_2$. 
The depicted curves increase with decreasing delay and reach the same value in the overlap ($\Delta t=0$). For a non-zero delay, the curve splits into three branches for $A_1>A_2$, $A_1=A_2$ and $A_1<A_2$.
%
%
To obtain the switching signal, we assumed the sigmoidal relation between  
$\Delta R$ and the maximal reached temperature, $T_{\mathrm{max}}$: 
\begin{equation}
\Delta R =  \frac{A_{\mathrm{S}}}{1+\exp(- \frac{T_{\mathrm{max}}-T_{\mathrm{C}}}{K} ) }
    \label{eq3}
\end{equation}
where $A_{\mathrm{S}}$ is the sigmoid amplitude,  $K$ is parameter characterizing its width and $T_{\mathrm{C}}$ sigmoid's center. The sigmoid function provides a good fit for the experimental
data (see inset of Fig.~1f) 
and represents the nonlinear activation function widely used in machine learning and neural networks. 

Fig.~\ref{figS5}d shows the resulting switching signal as a function of the pulse delay obtained by applying the sigmoidal function to the dependences in Fig.~\ref{figS5}b. 
The splitting of the switching signal into three separate branches for non-zero $\Delta t$ and 
different pulse amplitude ratios here is thus the consequence of the different maximal temperatures reached. Due to the application of non-linear sigmoid, the dependence of $\Delta R$ on $\Delta t$ is steeper than the  $T_{\mathrm{max}}$ vs $\Delta t$ dependence. From an experimental viewpoint, the difference between $T$ vs $\Delta t$ and the $\Delta R$ vs $\Delta t$ dependence is also the in the detection: Whereas the temperature is subject of ultrafast dynamics and appropriate method has to be chosen for the fast and precise sub-ns detection, the switching signal $\Delta R$ is governed by millisecond dynamics and can be recorded by a conventional electrical readout much later after the pulse impact. This way, we can obtain the information about the temperature through $\Delta R$. For the given $\Delta t$ and non-equal pulse amplitudes $A_1$ and $A_2$, it means we can determine their order thanks to the different resulting signal $\Delta R$.

\section{Applications using two unequal pulses}
\label{supp_sec3:appl}

\subsection{Recognition of pulse patterns based on experimental data}
\begin{figure}[h!]
\centering
\includegraphics[width=\textwidth]{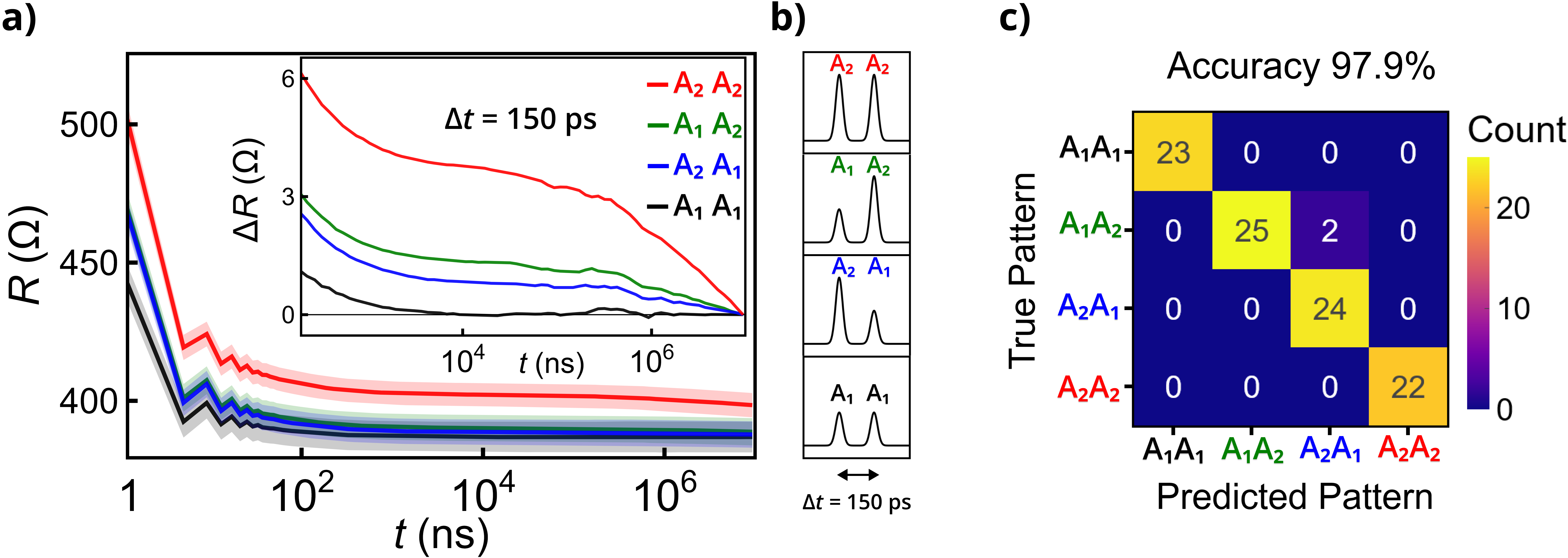}
\caption{\textbf{Recognition of temporal pulse patterns using resistance readout.} 
\textbf{(a)}~Experimental resistance data in the 10-ms range for all four combinations of pulse amplitudes $A_1=11.15$~mJ/cm$^2$ and $A_2=22.30$~mJ/cm$^2$ (pictured in \textbf{(b)}) and pulse delay $\Delta t = 150$~ps. For each of the four combinations, the dataset includes 48 resistance traces represented by their mean value (thick line) and 
standard deviation (shaded region corresponds to $\pm 0.5$~$\sigma$), totalling 192 traces. The inset shows the detail of $\Delta R$ from 200~ns to 10~ms. \textbf{(c)}~Results of the pulse pattern recognition with the test accuracy of $97.9~\%$ displayed as a confusion matrix. Further details in the text.
}
\label{figS3suppl}
\end{figure}

To determine pulse patterns of two pulses at sub-ns times from noisy experimental
data (Fig.~\ref{figS3suppl}a), we performed the classification into four classes corresponding to the four possible combinations of pulses with amplitudes $A_1$ and $A_2$ using the support vector machine classifier SVC from Python scikit-learn package \cite{pedregosa2011scikit}. The input dataset consisted of 192 class-balanced resistance traces with 100 log-spaced data points complemented by the size of the switching signal for each of the traces. We splitted the dataset equally into train and test subsets and performed the supervised machine learning on the train data. To find a
proper SVC classifier model and its hyperparameters, we employed the GridSearchCV routine from the scikit-learn package, examining 1200 models in total with various support vector machine kernels ('linear', 'poly', 'rbf' or 'sigmoid') and values of a regularizing parameter $C$. The best result, which lead to test accuracy of $97.9 \%$, was obtained for the polynomial kernel of fourth degree, and the value of $C = 0.001149$. Reducing the dataset traces to data points ranging from 200~ns to 10~ms where the switching signal dominates the resistance variation and repeating the same procedure lead to somewhat lower accuracy of $~86\%$.

\subsection{Convolutional edge detection}

\begin{figure}[ht]
 \centering
 \includegraphics[width=\textwidth]{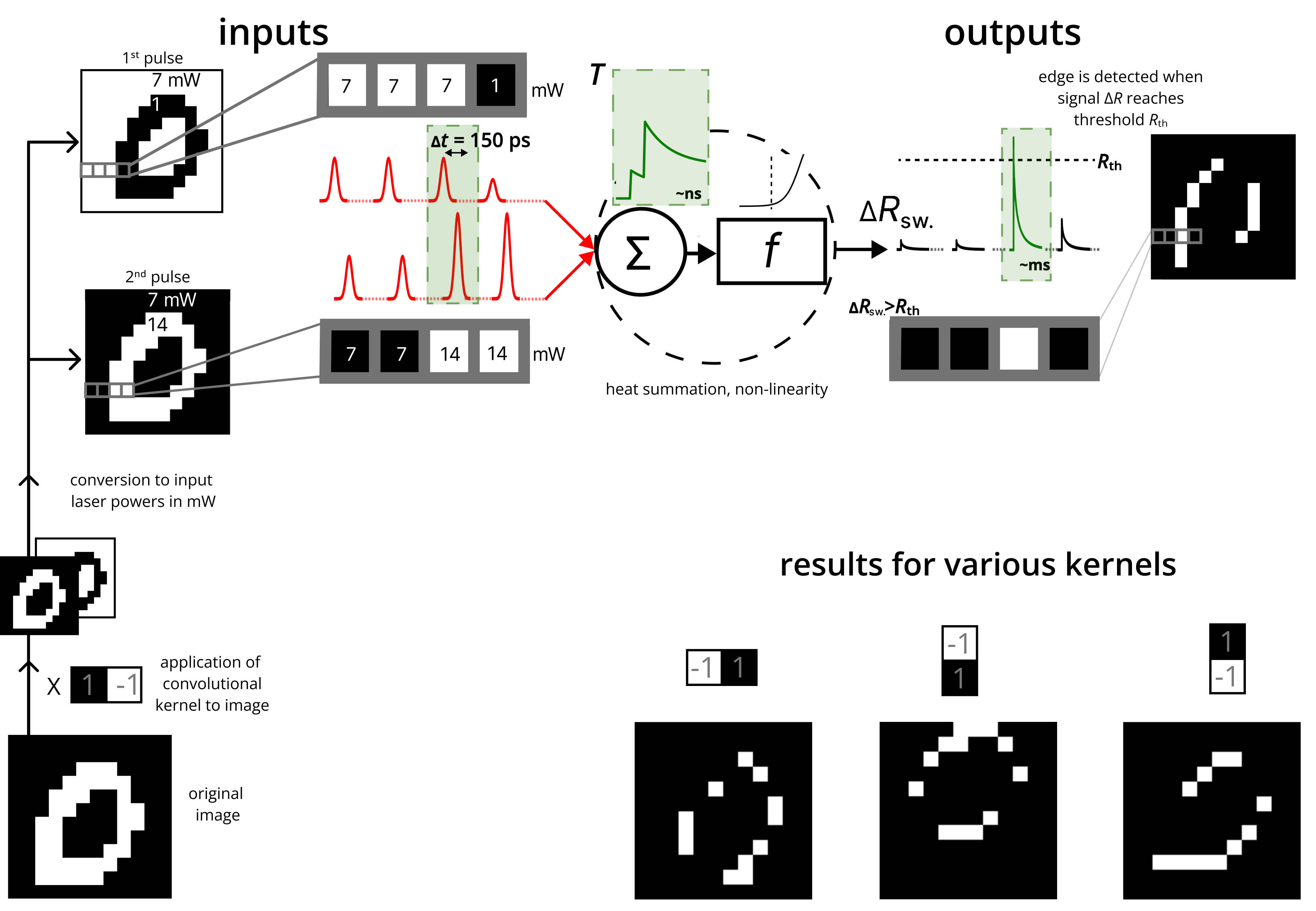}
 \caption{ \textbf{Recognizing patterns of pulses used for convolutional edge detection.}
 Application of the 2$\times$1 convolutional kernel to the original binary image leads to a Hadamard product of the size twice as large as the original image which corresponds to the copy of the original image and its inverse shifted by one pixel. The binary values of the image are converted to the laser fluences for the individual pulses and shined pulse by pulse to the sample with a fixed delay of $\Delta t = 150$~ps. This leads accumulation of heat at ns-scale which determines the resistance signal $\Delta R$ measured at ms-scale. The edge is detected only when the resistance signal $\Delta R$ exceeds a certain limit which corresponds to the situation when the 7~mW pulse is followed by the 14~mW. Below are displayed results of the convolutional edge detection for various kernels.
} 
 \label{figS2}
\end{figure}

Next, in Fig.~\ref{figS2}, we propose the application of unequal pulse-pattern recognition for convolutional edge detection. We conveniently map the binary image pixel values onto the fluences of the two pulses. The pulses, delayed by $\sim 150$~ps, are then used as the experimental inputs - pulse amplitudes. The image is processed sequentially, similarly as in the case of Fig.~2. 
The detected edge corresponds to the situation when the stronger pulse follows the weaker and the switching signal exceeds a switching threshold $R_{\mathrm{th}}$: $\Delta R_{\mathrm{sw}, A_1<A_2} < R_{\mathrm{th}} < \Delta R_{\mathrm{sw}, A_1>A_2}$.

\newpage

\section{Experimental setup}
\label{supp_sec3:experimental}
The schematics of the experimental setup consisting of the optical part used for excitation of the sample and the electrical part for resistance readout is depicted in Fig.~\ref{figS4}.

\subsection{Electrical measurements}

Time-resolved electrical measurements were performed using Rohde \& Schwarz RTP064 high-performance oscilloscope with 6~GHz bandwidth for readout and Rigol DG1000Z as a voltage source. We recorded the voltage $U$ of the input resistance  of the oscilloscope connected in series with the sample. The sample's resistance was then obtained from
$R = \frac{U_{\mathrm{source}}*R_{\mathrm{scope}}}{U} - R_{\mathrm{scope}}$ by reducing $R$ with the oscilloscope attenuation factor: $R_{\mathrm{sample}}=R/\sqrt{10^{atten\mathrm{[dB]}/10}}$.  We typically collected 2.5M points over a 10~$\mu$s or a 10~ms window during one acquisition, which was started by receiving a timing reference signal from the laser.
To reduce the noise at high frequencies without losing the time resolution, we acquired the data in a stroboscopic regime with a 100-Hz laser duty cycle and averaged out curves from $\sim$100 acquisitions.

\subsection{Optical measurements}
Fig.~S4 shows the experimental setup consisting of the electrical readout and optical part, enabling the sample's excitation by a pair of time-delayed laser pulses. The optical part of the setup involves femtosecond laser system Pharos by Light Conversion  delivering pulses with duration of 150~fs and wavelength of 1030~nm, and the delay line (DL) allowing for adjusting the mutual delay of laser pulses ($\Delta t$) 
in the interval from -0.5 to 4~ns. Further details about the experimental setup can be found in Ref.~\cite{surynek2024picosecond}.


\begin{figure}[ht]
\centering
\includegraphics[width=\textwidth]{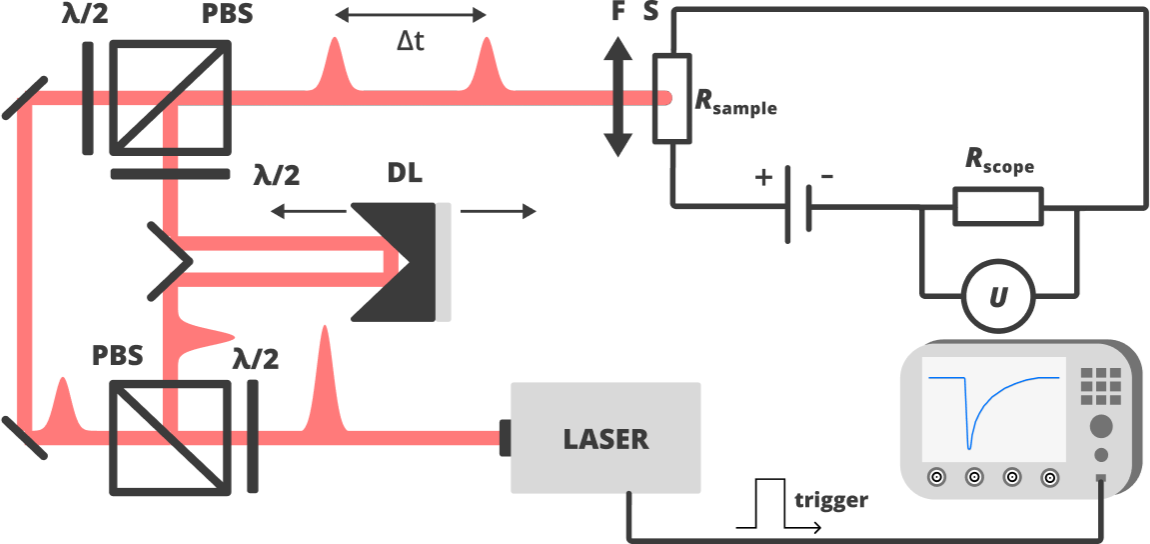}
\caption{ Experimental setup - optical (left) and electrical (right) part.
} 
\label{figS4}
\end{figure}


\subsection{Comparison of electrical and optical readout}
\begin{figure}[ht]
\centering
\includegraphics[width=\textwidth]{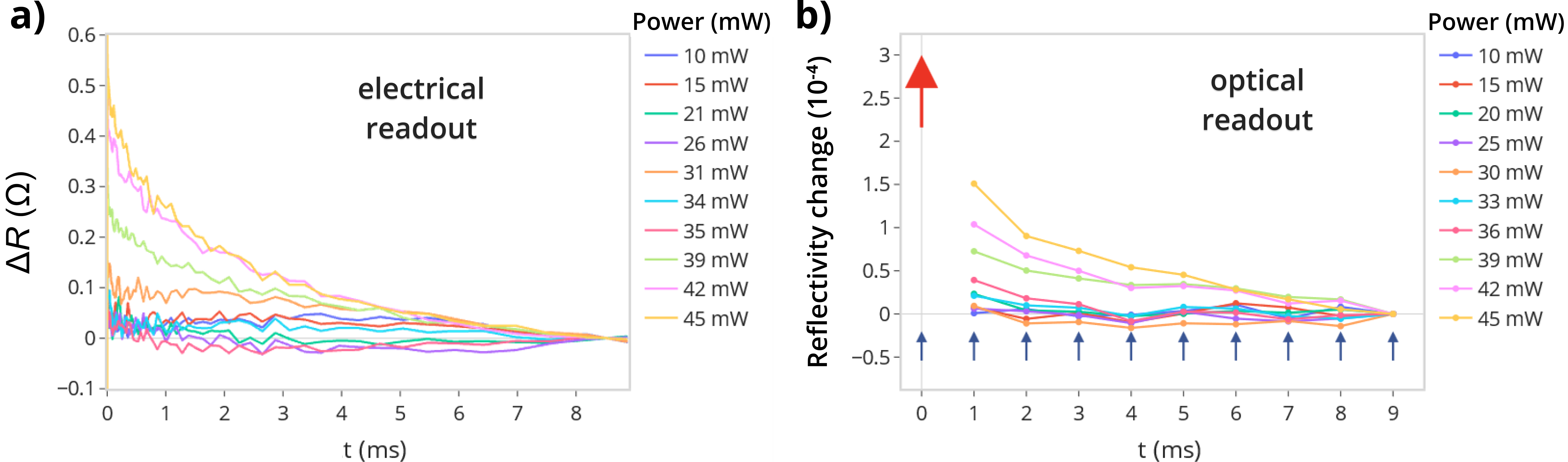}
\caption{ Comparison of electrical resistance readout \textbf{(a)} and the optical reflectivity readout \textbf{(b)} for several different laser powers. 
}
\label{figS6}
\end{figure}

Fig.~\ref{figS6} was obtained using the experimental setup shown in Fig.~\ref{figS4}. In addition to electrical readout, differential optical reflectivity was measured using an avalanche photodiode in a pump-probe configuration. Sub-threshold laser pulses probed the sample at a frequency of 1000 Hz, while the pump pulses were pulse-picked by a factor of 10, delivering optical excitation every 10 ms. The time delay between pump and probe pulses was fixed at $\Delta t = 100$~ps. By comparing the electrical readout (Fig.~\ref{figS6}a) with the optical reflectivity (Fig.~\ref{figS6}b) during the 10 ms writing time window, we correlated the onset of CuMnAs switching in the sample’s resistance with changes in reflectivity when excited with over-threshold pump pulses.

\clearpage
\bibliographystyle{unsrt} 
\bibliography{mybibliography_spintronics} 